\documentclass[12pt]{article}
\usepackage[dvips]{graphicx}
\usepackage{amssymb}
\usepackage{amsmath}
\usepackage{empheq}
\usepackage{epsfig}
\usepackage{cite}

\numberwithin{equation}{section}
\numberwithin{table}{section}

\def\beq{\begin{equation}}
\def\eeq{\end{equation}}
\def\be{\begin{equation}}
\def\ee{\end{equation}}
\def\bea{\begin{eqnarray}}
\def\eea{\end{eqnarray}}

\def\cV{\mathcal{V}}

\def\cM{\mathcal{M}}
\def\cN{\mathcal{N}}

\def\vev#1{{\langle #1\rangle}}

\def\AdS{\textrm{AdS}}
\def\mm{m}
\DeclareRobustCommand{\SkipTocEntry}[4]{}
\RequirePackage{color}

\def\vev#1{{\langle #1\rangle}}

\newcommand\iu{\operatorname{i}}

\begin{document}
\begin{titlepage}
\begin{center}
\rightline{\small ZMP-HH/15-13}
\rightline{\small CERN-PH-TH/2015-121}
\rightline{\small ITP-UH-13/15}

\vskip 1cm

{\Large \bf
$\cN=4$ Supersymmetric $\bf{\AdS_5}$ Vacua}

\vskip2mm

{\Large \bf and their Moduli Spaces}

\vskip 1.2cm

{\bf  Jan Louis$^{a,b}$, Hagen Triendl$^{c}$ and Marco Zagermann$^{d}$}

\vskip 0.8cm

$^{a}${\em Fachbereich Physik der Universit\"at Hamburg, Luruper Chaussee 149, 22761 Hamburg, Germany}
\vskip 0.3cm

{}$^{b}${\em Zentrum f\"ur Mathematische Physik,
Universit\"at Hamburg,\\
Bundesstrasse 55, D-20146 Hamburg, Germany}
\vskip 0.3cm

$^c${\em Theory Division, Physics Department, CERN, CH-1211 Geneva 23, Switzerland}

\vskip 0.3cm
$^d${\em Institut f\"{u}r Theoretische Physik \\
\& Center for Quantum Engineering and Spacetime Research,\\
Leibniz Universit\"{a}t Hannover, Appelstrasse 2, D-30167 Hannover, Germany}

\vskip 0.3cm

{\tt jan.louis@desy.de, hagen.triendl@cern.ch, marco.zagermann@itp.uni-hannover.de}

\end{center}

\vskip 1cm

\begin{center} {\bf ABSTRACT } \end{center}

\noindent We classify the $\cN=4$ supersymmetric $\AdS_5$ backgrounds that
arise as solutions of  five-dimensional $\cN=4$ gauged supergravity.
We express our results in terms of the allowed embedding tensor components and
identify the structure of the associated gauge groups.
We show that
the moduli space of these AdS vacua is of the form ${\rm SU}(1,\mm)/({\rm U}(1)\times
 {\rm SU}(\mm))$ 
and discuss our results regarding holographically dual $\cN=2$ SCFTs and their conformal manifolds.

\vfill

July 2015

\end{titlepage}

\section{Introduction}
Supersymmetric Minkowski compactifications of string or M-theory
 on Ricci-flat spaces generically result in
effective field theories with a large number of perturbatively
flat directions of the scalar potential. The geometry of these
moduli spaces is well understood, often even beyond the classical level,
and exploring the mechanisms that lead to moduli stabilization in realistic
string backgrounds is an important
task of string phenomenology.

Much less, by contrast, is known about the structure of moduli spaces of
anti-de Sitter (AdS) vacua. While such AdS moduli might
be encountered in intermediate steps of moduli stabilization scenarios, e.g.\
prior to de Sitter ``uplifts'', they play an even more fundamental
role in the context of the
AdS/CFT correspondence, where they correspond
to exactly marginal operators  of the holographically dual
conformal field theory (CFT). The space of exactly marginal couplings
is known as the  conformal manifold, ${\cal C}$, of the CFT, and it
comes equipped with the Zamolodchikov metric \cite{Zamolodchikov}.
Therefore,
knowledge of AdS moduli spaces can  provide valuable information
about ${\cal C}$.
Within the AdS/CFT correspondence, the study of ${\cal C}$ started in
\cite{Aharony:2002hx,Kol:2002zt,Tachikawa:2005tq}.

A first step towards a better  understanding of general AdS moduli spaces of string compactifications
is the investigation of anti-de Sitter solutions of lower-dimensional supergravity theories. These moduli spaces form submanifolds of the
scalar field spaces, ${\cal M}$, of the corresponding supergravity theories and may depend on additional data such as the gauge couplings or other deformation parameters.
Uncovering the interrelations between these geometric structures
defines an interesting mathematical problem in its own right that is highly sensitive to the spacetime dimension and the amount of supersymmetry present.

In \cite{deAlwis:2013jaa,Louis:2014gxa}, the moduli spaces of AdS${}_4$ vacua that preserve all the
available supersymmetries of four-dimensional (4D), $\mathcal{N}=1,2,4$ supergravity were investigated. For $\mathcal{N}=1$
supergravity, it was found in \cite{deAlwis:2013jaa} that the moduli space
${\cal C}$ is a real submanifold of the original K\"{a}hler manifold
${\cal M}$
with at best half the dimension. For $\mathcal{N}=2$ supergravity, ${\cal C}$
is generically a product of a real submanifold of the 
special-K\"{a}hler geometry
of the vector multiplet sector and a K\"{a}hler submanifold of the quaternion K\"{a}hler space of the hypermultiplets \cite{deAlwis:2013jaa}. For $\mathcal{N}=4$
supergravity, on the other hand, the moduli space was found to be
trivial in that only isolated AdS backgrounds can exist \cite{Louis:2014gxa}.
Although 4D supergravity is expected to capture at best parts of the holographic dual of
a 3D SCFT, the above results are consistent with what is known on
conformal manifolds of 3D superconformal field theories
\cite{Chang:2010sg,Green:2010da,Intriligatoretal}.
Motivated by the results of \cite{Louis:2014gxa} and the fact that
$\mathcal{N}=2$ SCFTs in 4D are intensely studied,\footnote{For a recent review see \cite{Tachikawa:2015bga} and references therein.}
 %
 we investigate, in this paper, fully supersymmetric AdS${}_{5}$ vacua of 5D, $\mathcal{N}=4$ supergravity theories
(i.e.\ AdS backgrounds that preserve all of the 16 real
supercharges).\footnote{In \cite{Louis:2015mka} a similar analysis is performed
  for supersynmmetric $\AdS_7$ backgrounds of seven-dimensional
  half-maximal supergravities where, as in $D=4$, no supersymmetric
  moduli space exists. Correspondingly, it can be shown that on the dual
  SCFT side no supersymmetric exactly marginal operators exist \cite{Intriligatoretal,Louis:2015mka}.}

5D, $\mathcal{N}=4$ gauged supergravities were constructed in
\cite{Romans:1985ps,Awada:1985ep,Dall'Agata:2001vb,Schon:2006kz},
and several specific examples of fully supersymmetric AdS${}_5$ vacua have previously appeared
in the literature.
In \cite{Romans:1985ps}, for instance, pure $\mathcal{N}=4$ supergravity with a gauge group
${\rm SU}(2)\times {\rm U}(1)$ was constructed and shown to exhibit
a fully supersymmetric $\AdS_5$ background. In this case, two of the six graviphotons have
to be dualized to tensor fields, which carry charge
under the ${\rm U}(1)$ factor of the gauge group. In \cite{Dall'Agata:2001vb},
the coupling of $\mathcal{N}=4$ supergravity to vector (or dual tensor) multiplets
was studied and particular $\AdS_5$ backgrounds were found -- again
for the gauge group ${\rm SU}(2)\times {\rm U}(1)$.
From the AdS/CFT perspective, the necessity of this gauge group
was discussed in \cite{Corrado:2002wx} for orbifold compactifications of type IIB string theory
dual to
4D, $\mathcal{N}=2$ superconformal quiver gauge theories
\cite{Kachru:1998ys}.
The 5D candidate gauged supergravity theory of the $\mathbb{Z}_n$ orbifolds of the five-sphere was identified in  \cite{Corrado:2002wx}
to be  a specific $\mathcal{N}=4$ truncation of $\mathcal{N}=8$ supergravity with additional vector and tensor multiplets
from the twisted sectors.
A moduli space of the form
${\rm SU}(1,m)/({\rm U}(1)\times {\rm SU}(m))$ was implicitly identified in \cite{Corrado:2002wx} for these theories by looking at the set of
holographic RG-flows induced by certain mass deformations.

Using the most general gaugings \cite{Schon:2006kz} in terms of the embedding tensor formalism \cite{Nicolai:2000sc,deWit:2002vt,deWit:2004nw},
we determine here the general gauge group that can lead to an $\mathcal{N}=4$ AdS${}_5$ vacuum and identify the possible moduli spaces.
The most general gauge group turns out to be of the form $G={\rm U}(1)\times H$, where $H$ must contain an ${\rm SU}(2)$ subgroup gauged
by three vector fields from the supergravity multiplet, and the ${\rm U}(1)$ must act at least on two tensor fields from the supergravity multiplet.
The general moduli space of these theories is shown to be of the form\footnote{This resembles the result for two-dimensional (4,4) SCFTs that have ${\rm SO}(4,\mm)/({\rm SO}(4)\times
 {\rm SO}(\mm))$ as conformal manifold \cite{Seiberg:1988pf,Cecotti:1990kz}.}
\begin{equation} \label{eq:modulispace}
{\cal C}=\frac{{\rm SU}(1,\mm)}{{\rm U}(1)\times {\rm SU}(\mm)} \ .
\end{equation}

Our analysis is intrinsically five-dimensional and at the classical
level, so that within the AdS/CFT correspondence
one would generally expect it to capture only part of the full story.
A sufficient condition for
the validity of a purely five-dimensional analysis in AdS backgrounds
is when the five-dimensional fields form a consistent truncation
of the ten-dimensional theory, as is for example the case for the untwisted sector of 5-sphere orbifold compactifications \cite{Corrado:2002wx}.
Working with classical supergravity
means that
the moduli space given in \eqref{eq:modulispace} should a priori only hold
in the large-$N$ limit ($N$ being the number of colors in the dual SCFT).
For an {\rm SU}(2) gauge group, for instance, it has indeed been shown in
\cite{Baggio:2014sna,Baggio:2014ioa} that
the Zamolodchikov  metric has a more complicated form which
agrees with the metric on ${\cal C}$ given in \eqref{eq:modulispace}
only at leading order.
Ref.~\cite{Papadodimas:2009eu} showed that
the conformal manifold in any 4D, $\mathcal{N}=2$ SCFT
is a K\"ahler manifold which in addition obeys the relations of $tt^*$
geometry \cite{Cecotti:1991me}.
Finally, Ref.~\cite{Gerchkovitz:2014gta} established that
the corresponding K\"ahler potential is given by the sphere partition
function of the SCFT while Ref.~\cite{Gomis:2014woa} proved
K\"ahlerness of the metric using supersymmetric Ward identities.
As we will show, consistency of (\ref{eq:modulispace})
with the $tt^*$
geometry of \cite{Papadodimas:2009eu} imposes a
constraint on the leading and subleading
large-$N$ behaviour of the three-point functions
that appear in the OPE of exactly marginal operators.

The paper is organized as follows.
In Section~\ref{Prelim}, we recall the properties of $\mathcal{N}=4$ gauged supergravity
that we need for our analysis.
In Section~\ref{AdS}, we analyze $\mathcal{N}=4$ AdS$_5$ backgrounds
and determine the constraints on the embedding tensor.
We then show that an
$\textrm{SU}(2)\times \textrm{U}(1)$
group is necessarily gauged by the graviphotons, and we also determine the allowed structure
of the full gauge group $G$, thereby classifying all possible $\mathcal{N}=4$ AdS${}_5$ vacua.
In Section~\ref{moduli}, we determine the moduli space of the above
AdS vacua,
and in Section~\ref{sec:SCFT} we discuss our results in terms of  dual 4D, $\cN=2$ SCFT.
Finally, Appendix~\ref{N4} summarizes our $\Gamma$-matrix conventions,
while in Appendix~\ref{sec:largeN} we discuss the large-N
behaviour of correlation functions in the SCFT and the constraints
which can be derived from the consistency with (\ref{eq:modulispace}).


\section{$\mathcal{N}=4$ gauged supergravity}\label{Prelim}
\noindent In this section, we recall the properties of 5D, $\mathcal{N}=4$
gauged supergravity \cite{Romans:1985ps,Awada:1985ep,Dall'Agata:2001vb, Schon:2006kz} that are relevant
for our analysis. The generic spectrum of ungauged $\mathcal{N}=4$ supergravity
consists of the gravity multiplet together with $n$~vector
multiplets.  The gravity multiplet contains the graviton $g_{\mu\nu}$, four
gravitini $\psi^i_{\mu},\, i=1,\ldots,4$, six vectors  $A_\mu^{[ij]}, A_\mu^0$,
four spin-1/2 fermions $\chi^i$, and one real scalar $\Sigma$.
The vector fields $A_\mu^{[ij]}$ are antisymmetric in $i$ and $j$ and satisfy
the additional condition
\be\label{Aconstraint}
A_\mu^{[ij]}\Omega_{ij} =0\ ,
\ee
where $\Omega_{ij}$ is the symplectic metric of ${\rm USp}(4)$, the R-symmetry group of 5D, $\mathcal{N}=4$
supersymmetry.
Thus, the $A_\mu^{ij}$ transform in the ${\bf 5}$ of ${\rm USp}(4)$, while
$A_\mu^0$ is a ${\rm USp}(4)$ singlet.

We label the vector multiplets with the index $a=1,\ldots,n$. Each vector multiplet
contains a vector $A_{\mu}^a$, four spin-1/2 gaugini $\lambda^{ai}$,
and $5$ scalars $\phi^{a[ij]}$, which are also antisymmetric in $i$ and $j$ and symplectic traceless analogous to \eqref{Aconstraint}.
Altogether, the spectrum thus features the graviton,
four gravitini, $(6+n)$ vector bosons,
$(4+4n)$~spin-1/2~fermions, and $(5n+1)$ scalars.

The target space, $\cM$, of the scalar fields is the coset
\be\label{N4coset}
\cM = {\rm SO}(1,1)\times  \frac{{\rm SO}(5,n)}{{\rm SO}(5)\times {\rm SO}(n)}\ ,
\ee
where the first factor is spanned by $\Sigma$ while the second factor
is spanned by the scalars $\phi^{a[ij]}$ in the vector multiplets.

The second factor in \eqref{N4coset} is conveniently parametrized by
the vielbein $\cV =(\cV^{m}_M, \cV^a_M)$, with $M=1,\ldots,n+5, m=1,\ldots,5$.
$\cV$ is an element of ${\rm SO}(5,n)$ and thus obeys
\begin{equation} \label{eq:vielbein_metric}
 \eta_{MN} = - \cV^m_M \cV^m_N + \cV^a_M \cV^a_N \ ,
\end{equation}
where  $\eta_{MN}={\rm diag}(-1,-1,-1,-1,-1,+1,\dots,+1)$
is the flat ${\rm SO}(5,n)$ metric.
Alternatively, the coset can be represented by the positive definite scalar metric
\begin{equation}\label{eq:Mmetric}
M_{MN}\ =\ \cV^m_M \cV^m_N + \cV^a_M \cV^a_N \ =\ 2 \cV^m_M\cV^m_N + \eta_{MN} \ ,
\end{equation}
which also plays the role of the gauge kinetic matrix for the $(5+n)$
vector fields combined as $A_{\mu}^{M}=(A_{\mu}^{[ij]}, A_{\mu}^{a})$.

The isometry group of the scalar manifold, ${\rm SO}(1,1)\times {\rm SO}(5,n)$, extends to a global
symmetry of the entire ungauged supergravity action, which is also subject to a local composite
invariance under ${\rm Spin}(5)\times {\rm SO}(n)$. In order to express the boson-fermion couplings in a way that makes these symmetries manifest, one uses the group isomorphism
between ${\rm USp}(4)$ and ${\rm Spin}(5)$ to express the ${\rm SO}(5)$ index $m$ of the scalar vielbeine $\cV_{M}^{m}$ in terms of
${\rm USp}(4)$ indices $i,j$ via ${\rm SO}(5)$ gamma matrices,
\be
\cV^{ij}_M := \cV^m_M\, (\Gamma_m)^{ij}\ .
\ee
$\cV_M^{ij}$ is then antisymmetric and symplectic traceless
in $i$ and $j$ and hence transforms in the ${\bf 5}$ of ${\rm USp}(4)$. More details on this and our precise conventions are given in Appendix~\ref{N4}.

In the gauged versions of these theories, a subgroup of the global symmetry group ${\rm SO}(1,1)\times {\rm SO}(5,n)$
is promoted to a local gauge symmetry by introducing minimal couplings to the
gauge fields and a few further terms to restore supersymmetry.
This breaks part of the global symmetry group and, as a special feature
of five dimensions, may require the conversion of some of the vector fields
to antisymmetric tensor fields \cite{Romans:1985ps,Dall'Agata:2001vb}. This conversion concerns
vector fields that would transform in nontrivial representations of the gauge
group other than the adjoint representation
and
also occurs for $\mathcal{N}=8$ \cite{Gunaydin:1984qu,
Gunaydin:1985cu,Pernici:1985ju} and $\mathcal{N}=2$  \cite{Gunaydin:1999zx} supergravity.
In the case at hand, a conversion to tensor fields would in particular be necessary if the original
representation\footnote{The subscripts denote the charge under   ${\rm SO}(1,1)$. } $\mathbf{(5+n)_{-1}\oplus 1_{2}}$  of the global symmetry group ${\rm SO}(5,n)\times {\rm SO}(1,1)$
decomposes w.r.t. the gauge group $G\subset {\rm SO}(5,n)\times {\rm SO}(1,1)$ as
\begin{equation}
 \mathbf{(5+n)_{-1}\oplus 1_{2}}\longrightarrow \textrm{singlets of }G \oplus \textrm{non-singlets of } G \oplus \textrm{adj. of } G, \label{decomposition}
\end{equation}
and would then affect the non-singlets of $G$.

In the so-called embedding tensor formalism \cite{Nicolai:2000sc,deWit:2002vt,deWit:2004nw}, one can rewrite the theory such that the original global symmetry ${\rm SO}(1,1)\times {\rm SO}(5,n)$ remains manifest. In order to do this, one has to work with a redundant field content that contains a tensor field for each of the original vector fields. The gauge couplings are then described by three field-independent
${\rm SO}(1,1)\times {\rm SO}(5,n)$-tensors (the embedding tensors) denoted by
$\xi_{M},\xi_{[MN]}, f_{ [MNP]}$. Their transformation under ${\rm SO}(5,n)$ follows from the indicated index structure, and,
with respect to  ${\rm SO}(1,1)$, $\xi_{M}$ and $f_{[MNP]}$ carry  charge $-1/2$,
while $\xi_{[MN]}$ has charge $+1$.
The entries of the embedding tensors are real numbers, and
supersymmetry imposes
a set of coupled consistency conditions on them
known as the quadratic constraints \cite{Schon:2006kz}\footnote{Here and in the following, the ${\rm SO}(5,n)$ indices $M,N,\ldots$ are raised and lowered with $\eta^{MN}$ and $\eta_{MN}$ as in \cite{Schon:2006kz}. Consistency with $\cV_{M}^{A}\cV_{A}^{N}=\delta_{M}^{N}$ and $\cV_{A}^{M}\cV_{M}^{B}=\delta_{A}^{B}$ then requires raising and lowering
the ${\rm SO}(5)\times {\rm SO}(n)$ indices $A,B,\ldots$ with $\eta^{AB}$ and $\eta_{AB}$, i.e.\ we have
$\cV_{M}^{a}=\cV_{M a}$ and $\cV_{M}^{m}=-\cV_{M m}$. This differs from the conventions used in \cite{Awada:1985ep,Dall'Agata:2001vb}, where
$M,N,\ldots$  are raised and lowered with $M_{MN}$ and its inverse, while $A,B,\ldots$ are  raised and lowered
with the Kronecker delta to ensure consistency with $\cV _{B}^{N}$ being the inverse of $\cV_{M}^{A}$.}
\begin{equation}\label{eq:quadconstr}\begin{aligned}
 \xi^M \xi_{ M} =  0 \ ,& \qquad \xi_{MN}\xi^N=0\ ,\qquad
\xi^P f_{PMN} =  0 \ ,
 \\
 3 f_{ R[MN} f_{ PQ]}{}^R =& 2 f_{[MNP}\xi_{Q]}  \ ,
 \qquad
\xi_M{}^Qf_{QNP} = \xi_M\xi_{NP} - \xi_{[N}\xi_{P]M}\ .
\end{aligned} \end{equation}

The possible solutions to these constraints parameterize the different consistent gauged
$\mathcal{N}=4$ supergravity theories. In particular, they determine the gauge group and its precise embedding in the global symmetry group ${\rm SO}(1,1)\times {\rm SO}(5,n)$,
the order parameters for spontaneous supersymmetry breaking, and the scalar potential.

The full bosonic Lagrangian is recorded in  \cite{Schon:2006kz} but
for the analysis in this paper, we only need the  potential $V$
and the kinetic terms of the scalar fields, which are given by
\begin{equation}\begin{aligned}\label{Ldef}
  e^{-1}\mathcal{L}
\ =\
\tfrac{1}{16}(D_{\mu}M_{MN})(D^\mu M^{MN})
- \tfrac{3}{2}\Sigma^{-2}(D_{\mu}\Sigma)(D^\mu \Sigma)
- V(M,\xi,f)  + \ldots\ .
\end{aligned}\end{equation}
The gauge covariant derivative reads
\begin{equation}
  \label{s2:DMMN}
  D_{\mu}  = \nabla_{\mu} - A_\mu^M f_M{}^{NP} t_{NP} - A_\mu^0 \xi^{NP} t_{NP}
 - A_\mu^M \xi^{N} t_{MN} - A_\mu^M \xi_M t_{\hat 0} \, ,
\end{equation}
where $t_{MN}=t_{[MN]}$ are generators of ${\rm SO}(5,n)$, $t_{\hat 0}$ is the generator of ${\rm SO}(1,1)$, and we have absorbed
the gauge coupling into the embedding tensor components.

The conditions for a supersymmetric AdS-background
can be concisely  formulated in terms of
the scalar components of the
$\mathcal{N}=4$ supersymmetry transformations.
For the four gravitini $\psi^i_\mu$,
the four spin-1/2
fermions in the gravitational multiplet $\chi^i$, and the gaugini $\lambda^i_a$,
they are given by \cite{Schon:2006kz}
\begin{equation}\begin{aligned}\label{susytrans}
\delta \psi_{\mu i} = &\  D_\mu \epsilon_i + \tfrac{\iu}{\sqrt 6} \Omega_{ij}A_1^{jk} \Gamma_\mu
\epsilon_k + \dots \ , \\
\delta \chi_i = &\  \sqrt 2 \Omega_{ij}A^{kj}_2 \epsilon_k + \dots \ , \\
\delta \lambda^a_i = &\ \sqrt 2 \Omega_{ij} A_{2}^{a\, kj} \epsilon_k + \dots \ ,
\end{aligned} \end{equation}
where $\epsilon_j$ are the four supersymmetry parameters, and
the dots indicate terms that vanish in a maximally symmetric space-time
background.
The fermion shift matrices in these expressions are defined as
\begin{equation}\begin{aligned}\label{shiftM}
A^{ij}_1 \ = \ & \tfrac1{\sqrt 6}(-\zeta^{(ij)} + 2 \rho^{(ij)}) \ ,
\\
A^{ij}_2 \ = \ & \tfrac1{\sqrt 6}(\zeta^{(ij)} +  \rho^{(ij)} + \tfrac32\tau^{[ij]}) \ , \\
A_{2}^{a\, ij} \ = \ & \tfrac12 (-\zeta^{a[ij]} +  \rho^{a(ij)}
- \tfrac{\sqrt 2}4\tau^{a}\Omega^{ij}) \ ,
\end{aligned} \end{equation}
where
\begin{equation}\begin{aligned}\label{shiftM2}
\tau^{[ij]} &= \Sigma^{-1} \cV^{ij}_M \xi^M\ , \qquad
\tau^{a} = \Sigma^{-1} \cV^{a}_M \xi^M\ , \\
\zeta^{(ij)} &= \sqrt 2\Sigma^{2} \Omega_{kl }\cV^{ik}_M\cV^{jl}_N  \xi^{MN}\ , \qquad
\zeta^{a[ij]} = \Sigma^{2} \cV^{a}_M\cV^{ij}_N  \xi^{MN}\ ,  \\
\rho^{(ij)} &= -\tfrac23 \Sigma^{-1} \cV^{ik}_M\cV^{jl}_N \cV^P_{kl} f^{MN}{}_P\ , \qquad
\rho^{a(ij)} = \sqrt 2\Sigma^{-1} \Omega_{kl }\cV_M^a\cV^{ik}_N\cV^{jl}_P f^{MNP}\ .
\end{aligned} \end{equation}
In terms of the shift matrices, the scalar potential is given by
\begin{equation}
\label{s2:scalarpot}
\tfrac14 \Omega^{ij} V= \Omega_{kl}(A_{2}^{a \, ik} A_{2}^{a\, jl} + A_{2}^{ik} A^{jl}_2
- A_{1}^{ik}A_{1}^{jl})\ .
\end{equation}

\section{$\mathcal{N}=4$ AdS$_5$ backgrounds}\label{AdS}
In this section, we study $\mathcal{N}=4$ gauged supergravities
that admit a fully supersymmetric AdS$_5$ background, i.e.\ with all sixteen
supercharges left
unbroken. The latter requirement demands that the supersymmetry variations
\eqref{susytrans}
 have to vanish in the AdS$_5$ background.
Inspecting \eqref{susytrans} and \eqref{s2:scalarpot}, we see that this implies
\begin{eqnarray}
\vev{A_2^{ij}}= \vev{A_{2}^{a\, ij}} &=&0 \ , \label{susyshift1}\\
\vev{A_1^{ij}  A_{1kj}} &=& \frac{1}{4} |\mu|^2\,\delta_k^i\ , \label{susyshift2}
\end{eqnarray}
where $\vev{V} = -|\mu|^2$ is the cosmological constant, which arises from the covariant derivative in the gravitino variation, and $\vev{\cdot}$ indicates
that a quantity is evaluated in the AdS-background.

\subsection{Constraints on the gauging}
We will now extract the constraints that are imposed by (\ref{susyshift1}) and (\ref{susyshift2}) on the embedding tensor
components, i.e.\ on the possible gaugings that can lead to $\mathcal{N}=4$ AdS vacua.

Let us begin with the evaluation of  (\ref{susyshift1}).
Inspection of \eqref{shiftM} reveals that $A_{2}^{a\, ij}$
decomposes into three different representations of ${\rm USp}(4)$, so that
all three terms in  $A_{2}^{a\, ij}$ have to vanish separately in the vacuum.
Similarly, in $A_2^{ij}$ the last
term is antisymmetric and thus also has to vanish in the vacuum, while the first
two terms in $A_2^{ij}$ have to cancel each other.
Thus, eqs.\ (\ref{susyshift1}) are equivalent to
\be
\vev{\tau^{[ij]}}=\vev{\tau^{a}}=\vev{\zeta^{a[ij]}}=\vev{\rho^{a(ij)}}=0\ ,\qquad
\vev{\zeta^{(ij)}} + \vev{ \rho^{(ij)}}=0\ . \label{N4conditions1}
\ee
Using \eqref{shiftM2}, the vanishing of  $\vev{\tau^{[ij]}}$ and $\vev{\tau^{a}}$ immediately gives\footnote{We also see from \eqref{s2:DMMN} that
$D_\mu \Sigma$ depends only on $\xi_{M}$ and thus, for $\cN=4$ AdS backgrounds,
$\Sigma$ is uncharged and $D_\mu \Sigma$
reduces to an ordinary partial derivative.}
\begin{equation}
\xi^M=0\ .
\end{equation}
In order to evaluate the rest of (\ref{N4conditions1}), it is convenient to convert the ${\rm SO}(5,n)$ covariant  embedding tensor components $f^{MNP}$
and $\xi^{MN}$ to the ${\rm SO}(5)\times {\rm SO}(n)$ covariant tensors $f^{ABC}:= \vev{\cV_{M}^{A}}  \vev{\cV_{N}^{B}}  \vev{\cV_{P}^{C}} f^{MNP}$ and $\xi^{AB}:=  \vev{\cV_{M}^{A}}  \vev{\cV_{N}^{B}}  \xi^{MN}$. The splitting $A=(m,a)$ then defines components such as  $f^{mnp}$ or $\xi^{ma}$, i.e.\
\begin{equation}
f^{mnp}\equiv \vev{\mathcal{V}_M^m} \vev{\mathcal{V}_N^n}
\vev{\mathcal{V}_P^p} f^{MNP} \ , \qquad \xi^{ma} \equiv
\vev{\mathcal{V}_M^m} \vev{\mathcal{V}_N^a} \xi^{MN}
 , \qquad \textrm{ etc. }\ ,
\end{equation}
which is the way the embedding tensor appears in the
background values of the fermion shift matrices (\ref{shiftM2}).
We recall that the indices $a,b,\ldots$ and  $m,n,\ldots$ are raised
and lowered with, respectively, plus and minus the Kronecker delta.

Using this and the ${\rm SO}(5)$ $\gamma$-matrix notation of Appendix \ref{N4}, the remaining three equations of
(\ref{N4conditions1}) are now equivalent to
%
%
\begin{equation}\label{Asol}
\xi^{am}\Gamma_m =0\ , \qquad f^{amn}\Gamma_{mn}=0\ , \qquad
\frac{3}{\sqrt{2}}\vev{\Sigma^3}\,\xi^{mn} \Gamma_{mn}= - f^{mnp} \Gamma_{mnp}\ ,
\end{equation}
or, using (\ref{Gammaid1}) - (\ref{Gammaid3}),
\begin{equation} \label{relxif}
\xi^{am}=0\ , \qquad f^{amn}=0\ ,\qquad 3\sqrt{2} \vev{\Sigma^3}\, \xi_{qr}=
\epsilon_{mnpqr}f^{mnp} \ .
\end{equation}

It remains to analyze (\ref{susyshift2}).
Using 
the last equation in \eqref{N4conditions1},
it can be expressed solely
in terms of $\zeta^{(ij)}$ 
so that it becomes a constraint on $\xi^{mn}$:
\be
\frac{1}{4}|\mu|^2\,\mathbf{1}_4=-3\vev{\Sigma^4}\,\xi^{mn}\xi^{pq}\Gamma_{mn}\Gamma_{pq} =
-\frac{3}{2}\vev{\Sigma^4}\,\xi^{mn}\xi^{pq}\,\{ \Gamma_{mn}, \Gamma_{pq}\}.
\ee
With (\ref{Gammaid7}), this decomposes into the two conditions
\begin{eqnarray}
\xi^{[mn}\xi^{pq]} &=&0\ , \label{xixiGamma}\\
  \xi^{mn}\xi_{mn}&=&\frac{|\mu|^2}{24\vev{\Sigma^4}}\ \neq\  0\ .  \label{xiximu}
\end{eqnarray}

Note that for $\xi^{mn}=0$,  no $\mathcal{N}=4$ supersymmetric AdS$_5$ background can occur.
The condition $\xi^{mn}\neq 0$ means that
among the 5-plet of graviphotons of the ungauged theory
some are  necessarily charged under the ${\rm U}(1)$ gauge group, so that
these must be converted to antisymmetric tensor fields in order to
carry out the gauging, cf.\ our discussion around
(\ref{decomposition}). Interestingly, also in 4D, $\mathcal{N}=4$
gauged supergravity, an $\mathcal{N}=4$ AdS vacuum
requires a gauging with a special feature, namely magnetic gaugings \cite{Louis:2014gxa}.\footnote{In \cite{Gunaydin:2005bf}, the dimensional reduction of 5D, $\mathcal{N}=2$ supergravity with charged tensor fields to 4D was found to lead to magnetic gaugings in 4D.
This does not necessarily mean, however, that a dimensional reduction of the above 5D, $\mathcal{N}=4$ AdS vacua would yield the 4D, $\mathcal{N}=4$ vacua of  \cite{Louis:2014gxa}.}

Finally, inserting $\xi^{M}=0$
in \eqref{eq:quadconstr}, we see that the quadratic constraints
considerably simplify,
leaving only the Jacobi identity for the structure constants $f_{MN}{}^{P}$ and their orthogonality to $\xi^{MN}$:
\begin{eqnarray}
f_{ RM[N} f_{ PQ]}{}^R &=&0\ , \label{quadconAdSa}\\
\xi^{MQ} f_{QNP} &=& 0 \ .\label{quadconAdSb}
\end{eqnarray}

\subsection{Solving the constraints for $\xi^{MN}$ and $f^{MNP}$}
\label{sec:vacuum}
What is left to do is to solve the
constraints (\ref{relxif}), (\ref{xixiGamma}) --
\eqref{quadconAdSb}, which will then specify
 the possible gauge groups and their precise embeddings in ${\rm SO}(5,n)$.
These group structures become most transparent if one works with
the actual representation matrices of the gauge group as they appear in the gauge covariant derivative
(\ref{s2:DMMN}) (subject to $\xi^M=0$) when it acts on the scalar vielbein $\mathcal{V}_{M}^A$.
The latter transforms in the fundamental representation of ${\rm SO}(5,n)$,
where the generators $t_{MN}$ take the form
\begin{equation}
(t_{MN})_P{}^Q = \delta^Q_{[M}\eta_{N]P}\ , \label{generatordetail}
\end{equation}
so that the $\mathcal{V}_{M}^A$ couple to the gauge fields
$A_\mu^0$ and $A_{\mu}^{M}$ with, respectively, the representation matrices
\begin{eqnarray}
(T_{0})_{N}{}^P&:=& -\xi^{QR}(t_{QR})_N{}^{P}= \xi_{N}{}^{P}\ ,\label{T0}\\
(T_{M})_{N}{}^{P}&:=&- f_{M}{}^{QR}(t_{QR})_{N}{}^{P}= f_{MN}{}^{P}\ .\label{TM}
\end{eqnarray}
Eqs.\ \eqref{quadconAdSa} and \eqref{quadconAdSb} imply that these representation matrices satisfy the commutation relations
\begin{equation}
[T_{0},T_{M}]=0\ , \qquad [T_{M},T_{N}]=-f_{MN}{}^{P}T_{P}\ ,
\end{equation}
i.e.\ $T_{0}$ generates an Abelian group factor.
Let us now evaluate how  (\ref{relxif}), (\ref{xixiGamma}) and  (\ref{xiximu}) further constrain
$T_{0},T_{M}$ and their commutation relations.

We start with the equation $\xi^{ma}=0$, which implies that the ${\rm U}(1)$ factor
gauged by $A^0$ acts on the gravity multiplet
(via the generator $(T_{0})_m{}^{n}=\xi_{m}{}^{n}$) and on the vector multiplets
(via the generator $(T_{0})_a{}^b=  \xi_{a}{}^{b}$) independently, i.e.\ this ${\rm U}(1)$ is a subgroup of ${\rm SO}(5)\times {\rm SO}(n)$ in ${\rm SO}(5,n)$.

Next we consider the condition
$f^{mna}=0$. It implies that the $T_{m}$ close among themselves and hence generate a proper subgroup of the gauge group.
Moreover,  $(T_{m})_{M}{}^{N}=f_{mM}{}^{N}$ must be block diagonal so that this subgroup does not mix fields from the gravity multiplet with fields from the vector multiplets, i.e.\ it is a subgroup of ${\rm SO}(5)\times {\rm SO}(n)$.
We have thus found
 that $T_{0}$ and $T_{m}$ generate compact subgroups that do not mix gravity multiplet and vector multiplet sector, so that their action on these two sectors can be studied independently.

 We begin with the action of $T_0$ and $T_m$ within the gravity multiplet, which is described by the
 components $\xi^{mn}$ and $f_{mnp}$. Note that both of these tensors must be non-zero (and proprtional to the AdS
curvature), as eq.\
(\ref{xiximu}) requires $\xi^{mn}\neq 0$, which then also implies $f_{mnp}\neq 0$
by the last of eqs.\ (\ref{relxif}).
We now use that, by certain ${\rm SO}(5)$ transformations, the antisymmetric bilinear form $\xi^{mn}$ can always be brought to canonical form
where at most $\xi^{12}=-\xi^{21}$ and $\xi^{34}=-\xi^{43}$ are non-zero.\footnote{An ${\rm SO}(5)$ rotation about the 1-axis can rotate the vector $\xi^{1m}$ into the 2-direction, followed by a rotation about the 2-axis that rotates $\xi^{2m}$ along the 1-direction. Subsequent ${\rm SO}(3)$ rotations about the 4- and 3-axis can similarly eliminate all remaining components of $\xi^{3m}$ and $\xi^{4m}$ up to $\xi^{34}=-\xi^{43}$.}
Without loss of generality, we can assume $\xi^{12}=-\xi^{21}\neq 0$. The primitivity condition (\ref{xixiGamma}) then implies $\xi^{34}=-\xi^{43}=0$.

Since $\xi^{12}$ is the only nontrivial component of $\xi^{mn}$,
it implies, via (\ref{relxif}), that
the only non-vanishing structure constants $f_{mnp}$ are $f_{345}$ and
permutations thereof, so that the total gauge group that acts within the gravity multiplet is ${\rm U}(1)\times {\rm SU}(2)$.
Note that $\xi^{mn}f_{npq}=0$ (cf.\ (\ref{quadconAdSb})) is then automatically satisfied.
In the following, we split the index $m$ into $\tilde{m}=1,2$ and $m'=3,4,5$, so that
$\xi^{m'n'}=0=f^{\tilde{m}\tilde{n}\tilde{p}}$.

We now turn to the part of the gauge group that acts nontrivially on the vector multiplet sector, i.e.\
to the components $\xi^{ab}$, $f_{abm}$ and $f_{abc}$. Note that unlike $\xi^{mn}$ and $f_{mnp}$
none of these components necessarily needs
to be non-vanishing for an $\mathcal{N}=4$ supersymmetric AdS vacuum to exist.

We start with $\xi^{ab}$.
If $\xi^{ab}\neq 0$, we see from (\ref{T0}) that the ${\rm U}(1)$ gauged by
$A_{\mu}^0$ is a diagonal ${\rm U}(1)$ of a ${\rm U}(1)$ in ${\rm SO}(5)$ and a ${\rm U}(1)$ in ${\rm SO}(n)$, whereas for $\xi^{ab}=0$ it is entirely contained in ${\rm SO}(5)$.
Just as we did for $\xi^{mn}$, we can use suitable ${\rm SO}(n)$ transformations to bring also $\xi^{ab}$, and hence
 the ${\rm U}(1)$ generator $T_{0}$, into canonical block-diagonal form,
\begin{equation} \label{T0a}
T_{0}=\textrm{diag}(\alpha \epsilon,\textbf{0}_{3},\beta_1\epsilon,\beta_2\epsilon,\ldots,\beta_{p}\epsilon,0,\ldots,0)\ ,
\end{equation}
where $\alpha,\beta_1,\ldots, \beta_p$ are non-vanishing real numbers, which can always be assumed positive after possible exchanges of the relevant rows and coloumns, and $\epsilon=i\sigma_2$. Here, the special case $\xi^{ab}=0 $ is meant to corresponds to $p=0$, i.e.\ there would then be no $\epsilon$-blocks with $\beta$-coefficients.
In analogy with the above decomposition $m=(\tilde{m},m')$, we then decompose the indices $a,b,\ldots$ and use $\tilde{a},\tilde{b},\ldots = 1,\ldots,2p$ for
the directions in which $\xi^{ab}$ is non-trivial, and $a',b',\ldots=2p+1,\ldots,n$ for the rest.
The conditions $\xi^{\tilde{m}M}f_{MNP}=\xi^{\tilde{m}\tilde{n}}f_{\tilde{n}NP}=0$ and $\xi^{\tilde{a}M}f_{MNP}=\xi^{\tilde{a}\tilde{b}}f_{\tilde{b}NP}=0$ then imply that all components $f_{MNP}$ with at least one $\tilde{a}$ or one $\tilde{m}$
index must vanish, so that modulo index permutations only $f_{m'n'p'}$, $f_{a'b'm'}$ and $f_{a'b'c'}$ can be non-zero. The
$(5+n)\times (5+n)$-matrices $T_{m'},T_{a'}$ thus may have the following general form:



  \begin{equation}
T_{m'}=\left(\begin{array}{cccc}
\mathbf{0}_{2} & & &  \\
 & f_{m'n'}{}^{p'} & & \\
 & & \mathbf{0}_{2p} & \\
 & & & f_{m'a'}{}^{b'}
 \end{array}\right),\qquad  T_{a'}=  \left(\begin{array}{cccc}
\mathbf{0}_{2} & & &  \\
 & \mathbf{0}_{3} & & f_{a'm'}{}^{c'} \\
 & & \mathbf{0}_{2p} & \\
 & f_{a'b'}{}^{n'} & & f_{a'b'}{}^{c'}
 \end{array}\right). \label{generatorform}\end{equation}

Using the the above pattern of possibly nontrivial structure constants,
the Jacobi identity~(\ref{quadconAdSa}) implies
  that the three matrices $f_{m' a'}{}^{b'}$ form a representation of ${\rm SO}(3)$ on the vector multiplet sector,
\begin{equation} \label{repSO3}
f_{m' a'}{}^{b'}f_{n' b'}{}^{c'} - f_{n' a'}{}^{b'}f_{m' b'}{}^{c'}= -
f_{m'n'}{}^{p'} f_{p' a'}{}^{c'} \ ,
\end{equation}
or, equivalently, that the $T_{m'}$ as given in (\ref{generatorform}) satisfy the ${\rm SO}(3)$ algebra,
\begin{equation}
[T_{m'},T_{n'}]= - f_{m'n'}{}^{p'}T_{p'}\ ,  \label{commutationrules1}
\end{equation}
whereas the remaining commutators are of the form
\begin{equation}
[T_{m'},T_{a'}]= -f_{m'a'}{}^{b'}T_{b'}\ ,  \qquad
[T_{a'},T_{b'}]=-f_{a'b'}{}^{c'}T_{c'}-f_{a'b'}{}^{m'}T_{m'}\ .\label{commutationrules2}
\end{equation}

If $f_{m'a'b'}=0$, the gauge group, $G$, obviously simplifies to $G={\rm U}(1)\times {\rm SU}(2)\times H_{c}$,
where $H_{c}\subset {\rm SO}(n-2p) \subset {\rm SO}(n)$ is a compact subgroup with structure constants $f_{a'b'}{}^{c'}$ that only acts on the vector multiplets and
whose adjoint representation can be embedded into the fundamental representation of ${\rm SO}(n-2p)$.\footnote{Any semisimple compact group $H_{c}$ can be embedded in this way into an ${\rm SO}(N)$ for sufficiently large $N\geq \textrm{dim}(H_{c})$ by identifying the Cartan-Killing metric of $\textrm{Lie}(H_c)$ with the (relevant part of the) ${\rm SO}(N)$ metric.}

In the case $f_{m'a'b'}\neq 0$, the gauge group is instead given by
$G={\rm U}(1)\times H$, where $H\subset {\rm SO}(3,n-2p)\subset {\rm SO}(3,n)\subset {\rm SO}(5,n)$ must contain ${\rm SO}(3)$ as a subgroup
and is in general non-compact with commutation relations of the form
(\ref{commutationrules1})-(\ref{commutationrules2}). The simplest nontrivial example of this kind occurs for $n=3$ and  is given by $f_{m'n'p'}= -\epsilon_{m'n'p'}$, $f_{m'a'b'}= +\epsilon_{(m'-2)a'b'}$, and $f_{a'b'c'}=0$, i.e.~the $T_{m'}$ generate
${\rm SO}(3)$,
and the $T_{a'}$ generate three non-compact directions that transform as a triplet under the
${\rm SO}(3)$. Since their algebra closes again in the $T_{m'}$, the $T_{a'}$ and the $T_{m'}$ altogether generate the simple gauge group $H={\rm SO}(3,1)$.
By turning on $f_{a'b'c'}= \lambda \epsilon_{a'b'c'}$, the $T_{a'}$ get an admixture of a compact direction of the ${\rm SO}(3)$ acting on the vector multiplet sector. For $\lambda<2$, the gauge group remains ${\rm SO}(3,1)$. For $\lambda>2$, the gauge group becomes ${\rm SO}(3)\times {\rm SO}(3)$ instead. In the case of $\lambda =2 $, the gauge group becomes the non-semi-simple gauge group of Euclidean rotations and translations in three dimensions.

We should point out that in general for $H$ to be simple, one has to make sure that
the non-degenerate Cartan-Killing metric of $H$
can be embedded into the ${\rm SO}(3,n-2p)$ metric
$\textrm{diag}(---+\ldots +)$, with the negative entries corresponding to ${\rm SO}(3)\subset H$. This means that a simple $H$ must have ${\rm SO}(3)$ as its maximally compact subgroup. Similar to the 4D case \cite{deRoo:1985jh}, this severely restricts the possible simple gauge groups $H$ that can lead to $\mathcal{N}=4$ AdS vacua and leaves essentially the above $H={\rm SO}(3,1)$ and $H=SL(3,\mathbb{R})$ as the only possibilities.
For non-simple $H$ there are of course many more possibilities.

To summarize, the necessary gauge group structure for an $\mathcal{N}=4$ AdS${}_{5}$ vacuum is
\begin{equation}
G={\rm U}(1)\times H_{\rm nc}\times H_{\rm c} \ ,\label{Ggeneral}
\end{equation}
where $H_{\rm nc}$ has the ${\rm SU}(2)$ as its maximally compact subgroup that is gauged by three graviphotons, and $H_{\rm c}$ is a compact group that is gauged only under vector multiplet gauge fields.
The ${\rm U}(1)$ is a diagonal subgroup of a necessary ${\rm SO}(2)\subset {\rm SO}(5)$ and an optional
${\rm SO}(2)\subset {\rm SO}(n)$.
In the case of $H_{\rm nc}$ being simple we find that it is either ${\rm SO}(3)$, ${\rm SO}(3,1)$ or ${\rm SL}(3,\mathbb{R})$.



We finally note that all vector fields of the ungauged theory that are acted on non-trivially by $T_{0}$ must be dualized to antisymmetric tensor fields in the gauged theory, which is in particular true for $A_{\mu}^{1}$ and $A_{\mu}^{2}$ from the gravity multiplet.
This together with the gauge group ${\rm U}(1)\times {\rm SU}(2)$ in the pure supergravity sector is consistent with the fact that
the $\mathcal{N}=4$ $AdS_5$ superalgebra has R-symmetry group ${\rm U}(1)\times {\rm SU}(2)$ and that the gravity multiplet representing this R-symmetry group has four vector fields transforming as $\mathbf{3_0}\oplus \mathbf{1_0} $ and two antisymmetric tensor fields transforming as singlets under ${\rm SU}(2)$ and a doublet under ${\rm U}(1)$ (see e.g.\ \cite{Corrado:2002wx} for a related discussion).

\section{$\mathcal{N}=4$ moduli space}
\label{moduli}
In the previous section, we determined the general form of the gauge groups
that can lead to $\mathcal{N}=4$ supersymmetric AdS vacua.
The purpose of this section is to determine the $\mathcal{N}=4$ moduli spaces
of these vacua, i.e.\ the manifold of
 scalar field deformations that preserve all four
supersymmetries of a given $\mathcal{N}=4$ AdS background.
To this end, we use the same method as in \cite{deAlwis:2013jaa,Louis:2014gxa} and
 vary the
supersymmetry conditions \eqref{susyshift1}--\eqref{susyshift2} so as to find
all possible directions in the scalar field space~${\cal M}$
that are left undetermined when \eqref{susyshift1}--\eqref{susyshift2} are preserved.
More concretely, we look for continuous solutions of
\be
\delta A_1^{ij}= \delta A_2^{ij}= \delta A_{2a}^{ij} =0 \ ,\label{SUSYFlat}
\ee
in the vicinity of a fully supersymmetric AdS$_5$ background.\footnote{Note that the scalar potential is quadratic in $A_1^{ij}, A_2^{ij}, A_{2a}^{ij}$ so that the solutions of (\ref{SUSYFlat}) are automatically flat directions of the scalar potential.} To start with, we parameterize the variations of
the vielbein $\mathcal{V}$ by  defining the $5n$ scalar field fluctuations
$\delta \phi^{ma}$ around an AdS$_5$ background value $\langle \mathcal{V}\rangle$ by
\begin{equation}
\mathcal{V}=\langle \mathcal{V}\rangle \, \exp [2 \, \delta\phi^{ma}
(t_{ma})]\ ,
\end{equation}
where $t_{ma}$ are the $(5+n)\times (5+n)$ matrices given in (\ref{generatordetail}) corresponding
to the coset ${\rm SO}(5,n)/({\rm SO}(5)\times {\rm SO}(n))$. This implies
\begin{equation}
\delta {\cal V}^m_M =  \vev{{\cal V}^a_M}\, \delta \phi^{ma}\ , \qquad
\delta {\cal V}^{a}_M = \vev{{\cal V}^{m}_M}\, \delta \phi^{ma} \ ,\label{VarV}
\end{equation}
which are also consistent with \eqref{eq:vielbein_metric}.
For the inverse vielbein, consistency with the relation $\mathcal{V}_{A}{}^{M}\mathcal{V}_{M}{}^{B}= \delta_{A}{}^{B}$ gives
\begin{equation}\label{varnuinv}
\delta {\cal V}_m^M  = -\vev{{\cal V}^M_a} \,\delta \phi^{ma}\ , \qquad
\delta {\cal V}_{a}^M = - \vev{{\cal V}^M_{m}}\, \delta \phi^{ma}\ .
\end{equation}
To linear order in $\delta \phi$, the metric $M_{MN}$ defined in
\eqref{eq:Mmetric} is then given by
\begin{equation} \label{eq:scalarmatrix}
 M_{MN} = \vev{M_{MN}} + 4 \vev{{\cal V}^{m}_{(M}}  \vev{{\cal V}^a_{N)}} \delta \phi^{ma} + {\cal O}(\delta \phi^2) \ .
\end{equation}


Applying the above variations to the three equations \eqref{relxif} gives, respectively, the following
conditions on $\delta\phi^{ma}$  and $\delta \Sigma$:
\begin{eqnarray}
\xi^{nm}\delta\phi^{na} + \xi^{ab}\delta\phi^{mb}&=&0\ ,\label{eq:moduli1}\\
 f^{pmn}\delta\phi^{pa} + f^{abn}\delta\phi^{mb} + f^{amb}\delta\phi^{nb} &=&0\ ,\label{eq:moduli2}\\
\delta \Sigma &=&0\ , \label{eq:sigma}
\end{eqnarray}
where, for the last equation, we used the identities $\delta \xi^{mn}=0$ and $\delta f^{mnp}=0$. These are simple consequences of
(\ref{VarV}) and $\xi^{ma}=0=f^{mna}$, which, together with  (\ref{eq:sigma}), also imply that
 (\ref{xixiGamma}) and (\ref{xiximu}) are automatically preserved.

Thus (\ref{eq:sigma}) fixes $\Sigma$, while (\ref{eq:moduli1}) and (\ref{eq:moduli2}) are the only nontrivial conditions
 on the other moduli. We will now show that these conditions mean that the moduli space is
 isomorphic to the coset space
${\rm SU}(1,\mm)/({\rm U}(1)\times {\rm SU}(\mm))$ for some  $\mm\leq p$ where $p$ denotes
the index range for which $\xi^{\tilde{a}\tilde{b}}$ is nontrivial
(cf.\ the discussion in the previous section below \eqref{T0a}).

To see this, we first examine (\ref{eq:moduli1}). As only $\xi^{\tilde{m}\tilde{n}}$ and $\xi^{\tilde{a}\tilde{b}}$ can be
 non-vanishing, eq.~(\ref{eq:moduli1}) is trivial for $(m,a)=(m',a')$ and yields three nontrivial equations for the other index combinations:
\begin{eqnarray}
\delta \phi^{\tilde{n}a'}=0, \qquad \delta\phi^{m'\tilde{b}}=0 \ ,&&\\
\xi^{\tilde{n}\tilde{m}}\delta\phi^{\tilde{n}\tilde{a}} + \xi^{\tilde{a}\tilde{b}}\delta\phi^{\tilde{m}\tilde{b}}=0\ . &&\label{Bedingung2}
\end{eqnarray}
Thus, only $\delta \phi^{m'a'}$ and $\delta \phi^{\tilde{m}\tilde{a}}$ can be nontrivial, with the latter being constrained by
(\ref{Bedingung2}).

Eq. (\ref{eq:moduli2}), finally, only constrains the components $\delta\phi^{m'a'}$
to satisfy
\begin{equation}
f^{p'm'n'}\delta\phi^{p'a'} + f^{a'b'n'}\delta\phi^{m'b'} +
f^{a'm'b'}\delta\phi^{n'b'} =0\ .
\end{equation}
This constraint was
already discussed in detail in \cite{Louis:2014gxa}, where it was shown that its solution is given by
\begin{equation} \label{eq:moduli3}
 \delta \phi^{m'a'} =  f^{ a'b' m'} \lambda^{b'}\ ,
\end{equation}
where $\lambda^{b'}$ is an arbitrary (infinitesimal) real vector.

Eq. (\ref{eq:moduli3}) implies that $\delta\phi^{m'a'}$ can only be nontrivial for $f^{a'b'm'}\neq 0$, i.e.\ for
non-compact gauge groups. Moreover, if we consider $(X_{a'}{}^{b'm'}):=f_{a'}{}^{b'm'}$ as a $(q\times 3q)$ matrix
(where $a',b'\ldots = 1,\ldots, q$), we see that the number of independent
 $\delta \phi^{m'a'}$ is equal to $\textrm{rk}(X)\leq q$, which is also the number of independent non-compact gauge group generators. As the non-compact gauge symmetries have to be spontaneously broken in a given vacuum, the
 $\delta\phi^{m'a'}$ are the natural candidates for the Goldstone bosons eaten by the corresponding non-compact gauge fields.
The physical moduli space would then only consists of the scalars
 $\delta \phi^{\tilde{m}\tilde{a}}$ subject to the constraint (\ref{Bedingung2}). We now confirm explicitly that the $\delta\phi^{m'a'}$ are indeed the Goldstone bosons eaten by the massive vectors and then give the geometric interpretation of the constraint (\ref{Bedingung2}) to identify the physical moduli space.

In order to identify $\delta \phi^{m'a'}$ with  Goldstone bosons, we
 consider the gauge covariant derivative of the scalar field matrix $M_{MN}$
(cf.\ \eqref{eq:scalarmatrix}) and introduce 
$D_{\mu} M_{AB}:= \langle\mathcal{V}_{A}^M\rangle \langle
\mathcal{V}_{B}^{N}\rangle D_{\mu}M_{MN}$. Using \eqref{s2:DMMN}
and  keeping only the linear terms in $\delta \phi$ and $A_{\mu}^M$,
we obtain
\begin{equation}
D_{\mu}M_{AB}= \langle\mathcal{V}_{A}^M\rangle \langle \mathcal{V}_{B}^{N}\rangle
\Big(4\langle \mathcal{V}_{(M}^{m}\rangle \langle \mathcal{V}_{N)}^{a}\rangle \partial_{\mu}
\delta\phi^{ma} +2A_{\mu}^P f_{P(M}{}^{Q}\langle M_{N)Q}\rangle + 2 A_{\mu}^{0}\xi_{(M}{}^{Q}\langle M_{N)Q}\rangle +\ldots \Big)\ .
\end{equation}
Introducing $A_{\mu}^{C}:= \langle \mathcal{V}_{M}^{C} \rangle A_{\mu}^{M}$ and using $\langle \mathcal{V}_{A}^{M} \mathcal{V}_{B}^{N} M_{MN}\rangle =\delta_{AB}$, this can be written as
\begin{equation}
	D_{\mu}M_{AB}= 4\delta_{(A}^{m}\delta_{B)}^{a}\partial_{\mu}\delta \phi^{ma} + 2 A_{\mu}^{C}f_{C(A}{}^{D}\delta_{B)D} + 2A_{\mu}^{0}\xi_{(A}{}^{D}\delta_{B)D} + \ldots,
\end{equation}
 which for $(A,B)=(m',a')$ becomes, using (\ref{eq:moduli3}),
 \begin{equation}
 2f^{m'a'b'}\partial_{\mu}\lambda^{b'} -2A_{\mu}^{b'}f^{m'a'b'} +
 \ldots\ .
\end{equation}
From this expression, we read off that under a local gauge transformation $\delta A_{\mu}^{b'} =
\partial_{\mu}\Lambda^{b'} + \ldots$ with $\Lambda^{b'}=\lambda^{b'}$, the nontrivial flat directions $\delta \phi^{m'a'}$ are absorbed by the vector fields
$A_{\mu}^{b'}$. Moreover, we see that the kinetic term $D_{\mu}M_{MN}D^{\mu}M^{MN}=
D_{\mu}M_{AB}D^{\mu}M^{AB}$ in the action results in  mass terms of the form
$\hat M^2_{a' b'} \sim f_{a'}{}^{c'm'}f_{b'}{}^{c'm'}= (X X^{T})_{a'b'}$.
This precisely gives mass to the  ${\rm rk}(X)$ non-compact gauge bosons, which thus eat all independent $\delta \phi^{m'a'}$, as claimed above.
One also notes that in the $\mathcal{N}=4$ supersymmetric AdS-backgrounds
all four graviphotons $A^{m'}$,  $A^0$ remain massless and thus, as expected,
the ${\rm SU}(2)\times {\rm U}(1)$ part of the gauge symmetry is always unbroken.

We now return to the only true moduli, the $\delta \phi^{\tilde{m}\tilde{a}}$
that are subject to the constraint \eqref{Bedingung2}. For convenience we will assume the form \eqref{T0a} for $T_0$. In the following we show that, for $\beta_{i}=\alpha$ ($i=1,\ldots, p$), this constraint describes the canonical embedding of
\be
\frac{{\rm SU}(1,p)}{{\rm U}(1)\times {\rm SU}(p)} \ \subset \  \frac{{\rm SO}(2,2p)}{{\rm SO}(2)\times
{\rm SO}(2p)} \ \subset  \ \frac{{\rm SO}(5,n)}{{\rm SO}(5)\times {\rm SO}(n)}
\ ,\ee
 and hence that the $\mathcal{N}=4$ moduli space is isomorphic to ${\rm SU}(1,p)/({\rm U}(1)\times {\rm SU}(p))$.
If not all $\beta_{i}$ are equal to $\alpha$, the moduli space becomes ${\rm SU}(1,\mm)/({\rm U}(1)\times {\rm SU}(\mm))$ for some $\mm<p$.

To see this, we recall the canonical embedding of the Lie algebra $\mathfrak{su}(1,p)$ into the Lie algebra
$\mathfrak{so}(2,2p)$.
Obviously, $\delta \phi^{\tilde{m}\tilde{a}}$ parameterizes the coset space ${\rm SO}(2,2p)/({\rm SO}(2)\times {\rm SO}(2p))$.
Decomposing the $(2\times 2p)$ matrix $\delta \phi^{\tilde{m}\tilde{a}}$ into $(2\times 2)$ blocks $A_{i}$, $i=1,\ldots,p$,
\begin{equation}
\left(\delta\phi^{\tilde{m}\tilde{a}}\right)=\left(\begin{array}{ccc}
A_1 & \cdots & A_p
\end{array}\right),
\end{equation}
 the condition (\ref{Bedingung2}) becomes
 \begin{equation}
 \alpha \epsilon A_i - \beta_{i} A_i\epsilon =0 \quad \textrm{(no
   sum)}\ .
 \end{equation}
If $\alpha=\beta_{i}$, this implies $A_i= x_i \mathbf{1}_{2}+y_{i} \epsilon$ for some real numbers $x_i,y_i$, whereas
$\alpha\neq \beta_{i}$ implies $A_{i}=0$. Assuming $\alpha=\beta_i$
for all $i=1,\ldots,p$,
the $\mathfrak{so}(2,2p)$ matrix parameterized by the $\delta\phi^{\tilde{m}\tilde{a}}$,
\begin{equation}
\left(\begin{array}{cccc}
\mathbf{0}_{2} &A_1 & \cdots &A_p\\
A_1^{T} & \mathbf{0}_{2} & \cdots & \mathbf{0}_{2}\\
\vdots & \vdots & & \vdots\\
A_p^T & \mathbf{0}_2 & \cdots & \mathbf{0}_{2}
\end{array}\right)
\end{equation}
is thus equivalent to the non-compact part of a general $\mathfrak{su}(1,p)$ matrix,
\begin{equation}
\left(\begin{array}{cccc}
0 & x_1+iy_1  & \cdots &x_p+iy_p\\
x_1-iy_1 & 0 & \cdots & 0\\
\vdots & \vdots & & \vdots\\
x_p-iy_p & 0 & \cdots & 0
\end{array}\right)
\end{equation}
upon the canonical embedding $x+iy \rightarrow x\mathbf{1}_{2}+y\epsilon$ of $\mathbb{C}$ into  $\textrm{Mat}(2,\mathbb{R})$.
Now the starting point of our considerations was an arbitrary $\cN=4$ vacuum point. This means that condition \eqref{Bedingung2} holds not
only at the point of consideration, but also in a neighborhood in the
space of $\mathcal{N}=4$ vacua. Therefore the moduli space is
homogeneous and is given by exponentiating the modes fulfilling
\eqref{Bedingung2}. For $\beta_{i}=\alpha\ \forall i$, the scalars $\delta \phi^{\tilde{m}\tilde{a}}$ thus parameterize ${\rm SU}(1,p)/({\rm U}(1)\times {\rm SU}(p))$.

If some of the $\beta_i$ are not equal to $\alpha$, the corresponding $x_i$ and $y_i$ vanish and the moduli space is
${\rm SU}(1,\mm)/(({\rm U}(1)\times {\rm SU}(\mm))$ for $\mm<p$, where $\mm$ counts the number of $\beta_{i}$ that are equal to $\alpha$. This reduced moduli space is consistent with the fact that the
coefficients $\alpha$ and $\beta_{i}$ determine the charges and the masses of the tensor fields. Only for a particular mass of the tensor fields will there be a massless scalar in the corresponding tensor multiplet, which just corresponds to the case $\beta_i=\alpha$ for the relevant index $i$.

To summarize, the moduli space of an $\mathcal{N}=4$ supersymmetric
AdS${}_{5}$ vacuum is always of the form
\begin{equation}\label{eq:modulispace1}
{\cal C}\ =\	\frac{{\rm SU}(1,\mm)}{{\rm U}(1)\times {\rm SU}(\mm)}
\end{equation}
for some $\mm$ with $2\mm\leq n$, where $n$ denotes the original number of vector multiplets in the ungauged theory. In addition,
$\mm$ counts the number of tensor fields in tensor multiplets that are charged with respect to the ${\rm U}(1)$ gauge group factor with the same charge as the two tensor fields from the gravity multiplet.

The above type of moduli space was also found in \cite{Corrado:2002wx} in a particular subset of
5D, $\mathcal{N}=4$ gauged supergravity theories that arise in type IIB compactifications
on orbifolds of $S^5$. Our results show that \emph{all}  $\mathcal{N}=4$ AdS vacua
of 5D gauged supergravity have this moduli space. Note that for $m=1$ this gives the familiar
moduli space of $\mathcal{N}=4$ super Yang-Mills theory with the
metric $g\sim (\tau-\bar\tau)^{-2}$, which also occurs in the untwisted
sector of the half-maximally supersymmetric 5-sphere orbifolds discussed in \cite{Corrado:2002wx}.

The coset space $\mathbb{C}H^{\mm}:={\rm SU}(1,\mm)/(({\rm U}(1)\times {\rm SU}(\mm))$ is sometimes called the complex (or Hermitian) hyperpolic space and  has several geometric properties that are also important for the rest of this paper. We first note that $\mathbb{C}H^\mm$ is the non-compact Riemannian symmetric space
dual\footnote{The dual of a symmetric space $G/H$ with Cartan decomposition $\textrm{Lie}(G)=\textrm{Lie}(H)\oplus \mathfrak{k}$ is the symmetric space
$G'/H$ with Cartan decomposition
$\textrm{Lie}(G')=\textrm{Lie}(H)\oplus i\mathfrak{k}$ (cf.\ \cite{Helgason}).
If $G/H$ is compact and has positive sectional curvature, then $G'/H$ is non-compact and has negative sectional curvature, and vice versa.}
to the complex projective space $\mathbb{C}P^\mm={\rm SU}(1+\mm)/({\rm U}(1)\times {\rm SU}(\mm))$ and that
it is a Hermitian symmetric space of complex dimension $\mm$ with isometry group ${\rm SU}(1,\mm)$.
Like all Hermitian symmetric spaces, $\mathbb{C}H^\mm$ is a K\"{a}hler manifold, and a form of the
K\"{a}hler potential that makes the ${\rm SU}(\mm)$ isometry subgroup manifest is
\begin{equation}\label{KCH}
K=-M^3\ln(1-z^{i}\bar{z}^{i})\ ,	
\end{equation}
where $z^{i}$ $(i=1,\ldots, \mm)$ are dimensionless
local complex coordinates on the
manifold. For future use we also included the dependence on the
five-dimensional Planck mass
$M$ which up to this point was chosen to be unity.\footnote{For $m=1$ there
 exists a coordinate transformation which puts $K$ into the form
$K=-M^3\ln(\tau-\bar\tau)$.}
Note that for dimensionless scalar fields the metric  and $K$ have
mass dimension three (in 5D) and indeed from \eqref{KCH} one finds
\begin{equation}
	g_{i\bar j}= M^3\left(\frac{\delta^{ij}}{(1-z^k\bar{z}^k)}+\frac{\bar{z}^i z^j}{(1-z^k\bar{z}^k)^2}\right) \ .
	\end{equation}
$\mathbb{C}H^\mm$ is also a special-K\"{a}hler manifold with holomorphic prepotential (see e.g.\
\cite{Sabra:1996xg} for further details on the special-K\"{a}hler geometry in various symplectic frames)\footnote{$\mathbb{C}H^\mm$ is a
special-K\"ahler manifold of the ``local'' type, i.e.\ one that could arise in the vector multiplet sector of 4D, $\mathcal{N}=2$ supergravity,
but not in rigid 4D, $\mathcal{N}=2$ supersymmetry. Such a  distinction could not be given for the AdS${}_4$ moduli spaces studied in \cite{deAlwis:2013jaa}.}

\begin{equation}
	F(X)=\frac{i}{2} X^{I}\eta_{IJ} X^{J}\ ,
\end{equation}
where $(X^{I})=(X^{0},X^{i}), I=0,1,\ldots, m$ are homogeneous special coordinates
related to the $z^i$ via  $X^{i}/X^{0}=z^i$, and $\eta_{IJ}=\textrm{diag}(+1,-1,\ldots,-1)$.
In general, the Riemann curvature tensor of special-K\"ahler manifolds
obeys \cite{Cremmer:1984hj}
\begin{equation} \label{Curvature}
R^l{}_{j \bar{m}k}= - M^6 g^{l \bar l}C_{\bar{l}\bar{m}\bar{k}}g^{\bar{k}n}C_{nkj}
+ M^{-3}(g_{\bar{m}j}\delta^l_k+
g_{\bar{m}k}\delta^l_j)\ ,
\end{equation}
where $C_{ijk}= e^{K/M^{3}} F_{ijk}$, with $F_{ijk}$ being the
third derivatives of the prepotential $F$.\footnote{Here we follow the
  conventions of \cite{Andrianopoli:1996cm}. Note that $C_{ijk}$ and $R^{l}{}_{j\bar{m}k}$ are
  dimensionless, so that with $g_{i\bar{j}}\sim M^3$
both sides of \eqref{Curvature} are in fact proportional to
  $M^3$.}
Since for the case at hand $F$ is quadratic, we have $C_{ijk}=0$ and thus
the Riemann tensor of $\mathbb{C}H^\mm$ obeys
\begin{equation}
R^l{}_{j\bar{m}k}=  M^{-3}(g_{\bar{m}j}\delta^l_k+
g_{\bar{m}k}\delta^l_j)\ ,
\label{Curvature1}
\end{equation}
This property of ${\cal C}$ is closely related to the $tt^*$-geometry
of
the dual SCFT,  as we discuss in Appendix~\ref{sec:largeN}.

\section{Holography and the $\mathcal{N}=2$ SCFT conformal manifold} \label{sec:SCFT}

So far our analysis has been entirely within  $5D$,  $\mathcal{N}=4$ gauged
supergravity. As we mentioned in the introduction, one of the
motivations
to study supersymmetric AdS-backgrounds comes from the relation
to holographically dual superconformal field theory (SCFT) within
the AdS/CFT correspondence.
For the case at hand, this would be a $4D,\, \mathcal{N}=2$ SCFT with
eight ordinary and eight superconformal supercharges.
The holographic dictionary between higher-dimensional type IIB backgrounds
of the form
AdS$_D\times Y_{10-D}$, where $Y_{10-D}$ is an appropriate compact
manifold,
has been discussed  in \cite{Kehagias:1998gn,Morrison:1998cs} and
reviewed, for example, in \cite{Polchinski:2010hw}. Here we only focussed
on the AdS$_D$ factor and did not consider any relation to solutions
of higher-dimensional supergravities or string theories.
It has not yet been firmly established which aspects are captured
by our lower-dimensional analysis.
However, for consistent truncations it is expected
that the lower-dimensional supergravity does give reliable
predictions for the dual SCFT in the large-N limit.
General consistent truncations to five-dimensional $\mathcal{N}=2$ and $\mathcal{N}=4$ gauged supergravities have been performed for instance in \cite{Buchel:2006gb,Gauntlett:2007ma,Cassani:2010uw,Gauntlett:2010vu,Cassani:2010na,Bena:2010pr}, but most of these truncations focus on gauged supergravities where the AdS${}_5$ vacuum is only $\mathcal{N}=2$ supersymmetric.  It would be interesting to find consistent truncations to five-dimensional supergravity for models with $\mathcal{N}=4$ vacua, as for instance the examples of \cite{Corrado:2002wx}, and to understand whether localized sources in the higher-dimensional theory can be included in such an analysis.

If a suitable consistent truncation to 5D, $\mathcal{N}=4$ supergravity exists, one might still wonder
whether there could be moduli among the modes one has truncated out, in particular among the infinite tower of Kaluza-Klein modes.
While the high masses of generic KK modes would usually prevent them from being moduli, in AdS spacetimes there could be a scalar in a KK-multiplet
that has mass zero even though the other members of the multiplet have
smaller and/or larger masses, as happens e.g.\ in the KK decomposition of type IIB supergravity
on the five-sphere \cite{Gunaydin:1984fk,Kim:1985ez}.
An exactly marginal operator, however, also has to be a singlet of the R-symmetry group of the SCFT, so that any AdS-modulus candidate among the truncated modes
would have to be 
neutral under the $SU(2)\times U(1)$ part of the 5D gauge group. If this group is realized geometrically in the
compactification space, a modulus in a KK multiplet would have to be inert under this geometric symmetry, which is typically not the case.

Keeping such issues in mind, let us now become a bit more specific and discuss possible interpretations of our result.
In Section \ref{AdS}, we found that
the AdS-backgrounds necessarily have an unbroken ${\rm U}(1)\times {\rm SU}(2)$
symmetry gauged by the graviphotons, which indeed
corresponds to the ${\rm U}(1)\times {\rm SU}(2)$ R-symmetry of the dual $\mathcal{N}=2$ SCFT.
The unbroken gauge factor $H_c\subset {\rm SO}(n)$
has to be related to an
unbroken flavour symmetry of the SCFT. We also found that
non-compact symmetries can be gauged, but they are always
spontaneously broken in the vacuum.

In Section~\ref{moduli},
we derived the coset space ${\rm SU}(1,\mm)/({\rm U}(1)\times {\rm SU}(\mm))$
as the moduli space of the AdS-backgrounds.
In the dual SCFT, this corresponds to the conformal manifold,
i.e.~the space of exactly marginal couplings $\varphi^i$
\cite{Kol:2002zt}.
They deform a given SCFT, $S^*$, as
\begin{equation}\label{deform}
 S[\varphi] = S^* + \sum_i \int   {\varphi^i}{O_i} \ ,
\end{equation}
where the $O_i$ denote the exactly marginal operators of $S^*$.\footnote{The notation $S^*$ is somewhat symbolic as we include the possibility of non-Lagrangian theories. Furthermore, the marginal operators ${O_i}$ we are interested in preserve all supercharges and thus have scaling dimension $\Delta=2$, are $R$-symmetry singlets
and form the highest components of their $\cN=2$ superfields.}
This deformation space
is endowed with a natural metric, the
Zamolodchikov metric given by
\begin{equation}
g_{ij}({\varphi}) = x^{2\Delta}\langle {O_i(x) O_j(0)}\rangle_{S[\varphi]}\ .
\end{equation}
The holographic dictionary states that in the large $N$-limit this metric
should agree with the metric on the moduli space of AdS-backgrounds.
In Section~\ref{moduli}, we derived such moduli spaces in 5D supergravity, and thus it is of interest to do a more detailed
comparison.

First of all there is the question to what extent
the Zamolodchikov metric is already constrained by supersymmetry.
Mimicking an argument first employed by N.~Seiberg in \cite{Seiberg:1988pf},
one can promote $\varphi^i$ to a background supermultiplet.
This in turn constrains the metric of this multiplet
to obey the properties imposed by the supersymmetry of the given SCFT.
For example in an $D=4,\, \mathcal{N}=1$ SCFT this argument
constrains
$g_{ij}({\varphi})$ to be a K\"ahler metric, which has indeed been shown
by other means in \cite{Asnin:2009xx}.

In 4D, $\mathcal{N}=2$ SCFT, the marginal operators $O_i$ reside
in conformal chiral multiplets with Weyl weight $w=2$, while the
deformation parameters $\varphi^i$ are members of chiral multiplets
with $w=0$. Unfortunately, the geometry of Weyl multiplets
with arbitrary Weyl weight is not known.\footnote{For
  higher-derivative couplings of the Weyl multiplet see, for example,
\cite{deWit:2010za,Butter:2013lta}.}
In \cite{Papadodimas:2009eu} it was shown that the metric on ${\cal C}$ is K\"ahler and
additionally obeys the $tt^*$-geometry \cite{Cecotti:1991me}. Moreover, the K\"ahler potential gives the sphere partition function, as has been shown by using localization techniques in \cite{Gerchkovitz:2014gta}
and supersymmetric Ward identities in \cite{Gomis:2014woa}.
The moduli space ${\cal C}=\mathbb{C}H^m$ we obtained in Section~\ref{moduli}
is both K\"ahler and obeys the $tt^*$-geometry, as discussed in Appendix \ref{sec:largeN}.
In fact, it is the specific special-K\"ahler manifold with a quadratic prepotential.
Of course, in our approach we only capture the large-$N$ limit
of the exact Zamolodchikov metric and therefore we are led to conjecture
that our result arises only in that limit. In Appendix
\ref{sec:largeN}, we discuss in more detail the large-$N$ limit in
view of \cite{Papadodimas:2009eu} and argue for a
specific subleading behaviour of the Zamolodchikov metric as well as
the (single and double trace) operators of dimension four in cases where our analysis applies.
Our result also suggests that the sphere partition function of suitable $D=4, \mathcal{N}=2$ SCFTs should simplify in the large-$N$ limit to agree with the exponential of the K\"ahler potential of~\eqref{eq:modulispace1}.


\section{Conclusion}

In this paper, we identified all five-dimensional, $\mathcal{N}=4$ gauged supergravity theories that allow for $\mathcal{N}=4$ AdS${}_{5}$ vacua and
determined the moduli spaces of these solutions. The requirement of a fully supersymmetric AdS vacuum constrains the gauge group of the supergravity theory to be of the general form
${\rm U}(1)\times H$, where $H$ must contain an ${\rm SU}(2)$ subgroup gauged by three graviphotons, and
the ${\rm U}(1)$ factor is gauged by another graviphoton and must (at least) act nontrivially
on two tensor fields in the gravity multiplet. The moduli space of the resulting vacua was
found to be the special-K\"{a}hler manifold ${\rm SU}(1,\mm)/({\rm U}(1)\times {\rm SU}(\mm))$, where $\mm$ counts the number of tensor fields from tensor multiplets with the same ${\rm U}(1)$ charge as the two tensor fields from the gravity multiplet.

We discussed this result in the context of
the AdS/CFT correspondence, where the holographic dual of the AdS moduli space is given by the conformal manifold of dual 4D,
$\mathcal{N}=2$ SCFTs. In cases where the truncation to five
dimensions captures all essential features of the ten-dimensional
theory this determines the large-$N$ behavior of the conformal
manifold
and via the result of \cite{Gerchkovitz:2014gta} also the large-$N$
behavior of the sphere partition function of the
SCFT. Comparison with the $tt^{\ast}$-like geometry found in
\cite{Papadodimas:2009eu} indicates that our result might constrain the large-N behavior of three-point functions that appear in the OPE of exactly marginal operators.


\section*{Acknowledgments}

The work of J.L.\ was supported by the German Science Foundation (DFG) under
the Collaborative Research Center (SFB) 676 Particles, Strings and the Early
Universe.
He also thanks the Theory Group at CERN
for its kind hospitality during the initial stage of  this work.
The work of M.Z. was supported by the German Research Foundation (DFG) within the Cluster of Excellence QUEST.

We have benefited from conversations and correspondence with Nikolay
Bobev,
Vicente Cort\'es, Bernard
de Wit, Murat
G\"{u}naydin, Ken Intriligator, Hans Jockers, Severin L\"ust, Thomas Mohaupt, Kyriakos Papadodimas, Yuji Tachikawa,
J\"{o}rg Teschner, Stefan Theisen, Alessandro Tomasiello, Antoine Van Proeyen and Daniel Waldram.

\newpage
\appendix
\noindent
{\bf\Large Appendix}
\section{SO(5) vs.\ USp(4) bases}\label{N4}

 The R-symmetry group of the  $\mathcal{N}=4$ Poincar\'{e}
 superalgebra in five space-time dimensions is given by
${\rm USp}(4)\equiv {\rm U}(4)\cap {\rm Sp}(4,\mathbb{C})$.
We denote the corresponding symplectic form
by $\Omega^{ij}$, $i,j=1,\ldots,4$, so that ${\rm USp}(4)$ is generated by Hermitian$(4\times 4)$-matrices ${\rm U}_{i}{}^{j}$ that satisfy $U^T \Omega + \Omega U =0$.
The fermions of $\mathcal{N}=4$ supergravity transform in the fundamental representation of
${\rm USp}(4)$. In order to describe their couplings to the scalar fields $(\mathcal{V}_{M}^{m},\mathcal{V}_{M}^{a})$
of the coset space ${\rm SO}(5,n)/{\rm SO}(5)\times {\rm SO}(n)$,
one converts the ${\rm SO}(5)$ index $m=1,\ldots,5$ to ${\rm USp}(4)$ indices $i,j$ using the group isomorphism
${\rm USp}(4) \cong {\rm Spin}(5)$ that follows from properties of the ${\rm SO}(5)$ Clifford algebra.
In the following, we briefly review some useful identities related to this isomorphism and match it to the supergravity
conventions used in this paper
(for further details see e.g.\ \cite{Kugo:1982bn,West:1998ey}).

The Clifford algebra in five Euclidean dimensions is represented by $(4\times 4)$ gamma matrices
$\Gamma_{m}$, $m,n,\ldots=1,\ldots,5$ satisfying
\begin{equation} \label{eq:gammamatrices_cc}
 \{ \Gamma_m, \Gamma_n\} = 2  \delta_{mn} {\bf 1}\quad
 \Longleftrightarrow \quad
\Gamma_{m\, i}{}^j \Gamma_{n\, j}{}^k + (m\leftrightarrow n)
 = 2 \delta_{mn} \delta_{i}^k\ .
\end{equation}
As in any odd dimension, there
are actually two equivalence classes of irreducible representations of (\ref{eq:gammamatrices_cc}).
They differ in how one defines the
the fifth gamma matrix in terms of the first four, which leaves a sign ambiguity:
$\Gamma_5=\pm \Gamma_1\Gamma_2\Gamma_3\Gamma_4$. Apart from
eqs.\ (\ref{Gammaid3})-(\ref{Gammaid6}), all equations in this appendix are insensitive to this sign choice.

5D rotational invariance requires the $\Gamma_m$ to be traceless, and for a
Euclidean Clifford algebra, they may also always be chosen to be Hermitian, as we will assume from now on:
\begin{equation}
 \Gamma_m=\Gamma_m^\dagger\ .
\end{equation}
For any representation of this type, there exists then a ``charge conjugation matrix'' $C$
with the following properties:
\begin{equation}
\Gamma_{m}^{T} = \Gamma_m^{\ast}=  C\Gamma_mC^{-1} \ ,\qquad
C=-C^{T}\ ,\qquad C^{\ast}=-C^{-1}\ . \label{Cmatrix}
\end{equation}
These relations imply, in particular, that $(C\Gamma_m)$ and $(C\Gamma_{mnpq})$ are antisymmetric, whereas
$(C\Gamma_{mn})$ and $(C\Gamma_{mnp})$ are symmetric matrices.
Due to its antisymmetry and invertibility, we can identify $C$ with a symplectic form $\Omega$  as follows:
\begin{equation}
\Omega^{ij}:= C^{ij}\ ,\qquad \Omega_{ij}:= C_{ji}=-C_{ij}\ .
\end{equation}
Here, $C^{ij}$ denote the entries of $C$, whereas $C_{ij}$ are meant to be the components of the inverse matrix
$C^{-1}$ so that $C^{ij}C_{jk}=\delta_k^i$, and hence $\Omega^{ij}\Omega_{kj}=\delta_k^i$.
$\Omega$ can then be used to raise and lower ${\rm USp}(4)$ indices
$i,j,\ldots$ according to the convention \cite{Awada:1985ep}
\begin{equation}
 V^i=\Omega^{ij}V_j\ , \qquad V_i=V^j\Omega_{ji}\ .
\end{equation}
We can then define
\begin{equation}
 \Gamma_m^{ij}:= \Omega^{ik}\Gamma_{m\, k}{}^j = (C\Gamma_m)^{ij}\ , \qquad \Gamma_{m\, ij}:= \Gamma_{m\, i}{}^k\Omega_{kj}=
 (\Gamma_m C^{-1\, T})_{ij}\ .
\end{equation}
$\Gamma_m^{ij}$ has the properties
\begin{equation}
 \Gamma_m^{ij}=-\Gamma_{m}^{ji}\ , \qquad
 \Gamma_{m}^{ij}\Omega_{ij}=0\ , \qquad
 (\Gamma_m^{ij})^{\ast}= \Omega_{ il} \,\Omega_{jk}\Gamma^{lk}_m\ ,
\end{equation}
where the first identity is just the antisymmetry of $(C\Gamma_m)$, the second is the tracelessness of
$\Gamma_m$, and the third equation a consequence of the reality properties
(\ref{Cmatrix}). Completely analogous identities are inherited by the coset representatives
\begin{equation}
\mathcal{V}_M^{ij}:=\mathcal{V}_M^{m}\Gamma_m^{ij}\ .
\end{equation}
Using the above properties, it is easy to see that the ${\rm SO}(5)$ generators
\begin{equation}
 M_{mn} := \frac{i}{4}\,[\Gamma_m,\Gamma_n]
\end{equation}
are Hermitian $(4\times 4)$-matrices that also satisfy
\begin{equation}
 (M_{mn})^T \cdot \Omega + \Omega \cdot M_{mn}=0\ ,
\end{equation}
i.e. that they can be viewed as generators of ${\rm USp}(4)$ in the fundamental representation.

We close with some useful identities:
\begin{eqnarray}
 \Gamma_m^{ij} \Gamma_{n\, ij} &=& 4\delta_{mn}\ , \label{Gammaid1}\\
 \textrm{tr} (\Gamma_{mn}\Gamma_{pq})&=&4(\delta_{mq}\delta_{np}-\delta_{mp}\delta_{nq}) \ , \label{Gammaid2}\\
 \Gamma_m&=& \pm\frac{1}{24}\epsilon_{mnpqr}\Gamma^{npqr}\ , \label{Gammaid3}\\
 \Gamma_{mn}&=& \mp \frac{1}{6} \epsilon_{mnpqr} \Gamma^{pqr}\ , \label{Gammaid4}\\
 \Gamma_{mnp}&=& \mp \frac{1}{2} \epsilon_{mnpqr}\Gamma^{qr}\ , \label{Gammaid5}\\
 \Gamma_{mnpq}&=& \pm \epsilon_{mnpqr}\Gamma^r\label{Gammaid6}\ , \\
 \{\Gamma_{mn},\Gamma_{pq}\}&=& 2\Gamma_{mnpq} +2\delta_{np}\delta_{mq}-2\delta_{nq}\delta_{mp}\ ,  \label{Gammaid7}
\end{eqnarray}
where we use $\epsilon_{12345}=1$, and the signs refer to the sign choice $\Gamma_5=\pm \Gamma_1\Gamma_2\Gamma_3\Gamma_4$.\footnote{In the main body of this work we will use the plus sign.}

\section{Large $N$ counting} \label{sec:largeN}
In \cite{Papadodimas:2009eu}, the Riemann tensor of the metric on the conformal manifold of a 4D,
$\mathcal{N}=2$ SCFT was found to satisfy the relation
\begin{equation}
R^{l}_{i\bar{j}k}=-C_{ik}^{M}g_{M\bar{N}}C_{\bar{j}\bar{q}}^{\ast \bar{N}} g^{\bar{q}l}
+g_{k\bar{j}}\delta_{i}^{l}+ g_{i\bar{j}}\delta_{k}^{l} \ .	\label{Riemann}
\end{equation}
Here, $C_{ij}^M$ are the chiral ring
coefficients between chiral primaries $O_{i}, O_{j}$ of conformal
dimension $\Delta=2$ and $O_{M}$ of conformal dimension $\Delta=4$, whereas $g_{i\bar{j}}$ and
$g_{M\bar{N}}$ denote the Zamolodchikov metrics for these operators.
The chiral ring coefficients can be expressed in terms of 3-point correlator coefficients
$C_{ij\bar{M}}$ as
\begin{equation}
	C_{ij\bar{M}}=C_{ij}^N g_{N\bar{M}}.	
\end{equation}
Note that all quantities in \eqref{Riemann} are dimensionless and no
powers
of any mass scale as in \eqref{Curvature} appear.

Our 5D supergravity analysis, on the other hand, led to AdS-moduli spaces of the form ${\rm SU}(m,1)/({\rm SU}(m)\times {\rm U}(1))$, which obeys \eqref{Curvature1}.
Since \eqref{Riemann} resembles \eqref{Curvature}, it is worthwhile
to establish a closer connection.
Note that the two formulas differ in that the OPE coefficients
$C_{ijM}$  do not coincide with the $C_{ijk}$ of
special geometry.  Therefore a comparison is not straightforward.
As the supergravity approximation in AdS/CFT is generally
only valid for large $N$ ($N$ being the number of colors), it
is useful to understand the large-$N$ behaviour
of the various terms in (\ref{Riemann}).

In \cite{Lee:1998bxa}, extremal 2- and 3-point correlators of single trace chiral primary
operators in 4D,
 $\mathcal{N}=4$ super Yang-Mills theories were computed in the weak
 coupling limit and at strong 't~Hooft
  coupling $\lambda=Ng_{YM}^{2}\gg 1$ using the dual supergravity side of the AdS/CFT correspondence. The results were found to agree.
  We recall here the $N$ dependence of the correlators in the weak coupling analysis.

We normalize
the  Yang-Mills action as $S=-\int\frac{1}{2g_{YM}^{2}}\mathbf{Tr}F^2+\ldots=-\int \frac{1}{4g_{YM}^{2}}F^a F^a+\ldots$,
where $F=F^a T^a$ with the ${\rm U}(N)$ generators $T^{a}$
$(a=1,\ldots,N^2)$, which we assume to be in the fundamental representation of ${\rm U}(N)$, i.e. they are $(N\times N)$ matrices with $\mathbf{Tr}(T^a T^b)=\frac{1}{2}\delta^{ab}$.

The scalar fields $\phi^\alpha=\phi^{\alpha}_{a}T^a$ $(\alpha=1,\ldots,6)$ have scaling dimension $\Delta =1$, transform in the fundamental representation of the R-symmetry group ${\rm SO}(6)$ and have the propagators
\begin{equation}
\langle \phi^{\alpha}_{a	}(x) \phi^{\beta}_{b} (y)\rangle
=\frac{g_{YM}^{2}\delta_{ab}\delta^{\alpha\beta}}{(2\pi)^2|x-y|^2}\ .
\end{equation}
As we are interested in massless supergravity scalar
 fields (the AdS moduli), we need to focus on marginal operators in
 the dual SCFT.  They have
scaling dimension $\Delta=2$ for the lowest component scalar
field (i.e.\ $\Delta =4$ for the highest component of the superfield) and
can be composed from two fundamental scalar fields $\phi_{a}^{\alpha}$ as a single trace operator\footnote{Here and in the following, the ${\rm SO}(6)$ indices $\alpha,\beta,\ldots$ should
always be thought of as being in  a completely symmetric and traceless combination, which, however, we do not make explicit as it does not affect the large $N$ scaling. Likewise, we are really interested in ${\rm SU}(N)$ instead of ${\rm U}(N)$ generators only.}
\begin{equation}
\mathcal{O}^{\alpha\beta}:=\mathbf{Tr}(\phi^{\alpha}\phi^{\beta})\ .
 \end{equation}
Using Wick's theorem, the free 2-point function  of two such single trace operators $\mathcal{O}$
is of the form
\begin{equation}
g(x,y)=\langle \mathcal{O}^{\alpha\beta}(x) \mathcal{O}^{\gamma\delta}(y)\rangle	= \frac{N^2 g_{YM}^4
(\delta^{\alpha\gamma}\delta^{\beta\delta}+\textrm{cyclic})}{(2\pi)^4|x-y|^4}\
.
\end{equation}
More generally, we have \cite{Lee:1998bxa}
\begin{equation}
g(x,y)=\langle \mathcal{O}^{\alpha_1\ldots\alpha_k} \mathcal{O}^{\beta_{1}\ldots\beta_{k}}\rangle	= \frac{N^k g_{YM}^{2k}
(\delta^{\alpha_1\beta_1}\ldots\delta^{\alpha_k\beta_{k}}+\textrm{cyclic})}{(2\pi)^{2k}|x-y|^{2k}}\
,
\end{equation}
for the single trace operators $\mathcal{O}^{\alpha_1\ldots\alpha_k}=
\mathbf{Tr}(\phi^{\alpha_1}\ldots\phi^{\alpha_k})$. We need the case
$k=4$ for the $\Delta= 4$ single trace operators, for which we read
off the scaling $N^4 g_{YM}^{8}$.

Next, let us consider
the 2-point function of the $\Delta= 4$ double trace operators
defined as $\mathcal{O}^{\alpha\beta,\gamma\delta}(x):=
\mathbf{Tr}(\phi^{\alpha}(x)\phi^{\beta}(x))\mathbf{Tr}(\phi^{\gamma}(x)\phi^{\delta}(x))$.
It scales like $N^4g_{YM}^8$, because Wick's theorem gives rise to terms such as
$\delta^{ab}\delta^{cd}\delta^{ef}\delta^{gh}\delta^{ae}\delta^{bf}\delta^{cg}\delta^{dh}\sim N^2 N^2$.
Note that among the dimension 4 operators that can be formed from the
scalars $\phi^\alpha$, there are only the single trace operators $\mathcal{O}^{\alpha\beta\gamma\delta}$ and the double trace operators $\mathcal{O}^{\alpha\beta ,\gamma\delta}$, when one restricts oneself to the traceless ${\rm SU}(N)$ generators.

The 3-point functions we need to consider are
thus of the form
\begin{equation}\begin{aligned}
&\langle \mathcal{O}^{\alpha\beta}\mathcal{O}^{\gamma\delta}	\mathcal{O}^{\epsilon\eta\kappa\lambda}\rangle\\
& \langle \mathcal{O}^{\alpha\beta}\mathcal{O}^{\gamma\delta}	\mathcal{O}^{\epsilon\eta,\kappa\lambda}\rangle.
\end{aligned}\end{equation}
The first 3-point function scales as $\lambda^4/N\sim N^3$ \cite{Lee:1998bxa}.
The second 3-point function can be directly determined with Wick's theorem and
gives a contribution that scales as $N^4g_{YM}^8$ (because it leads to $\delta_{ab}\delta^{ab}\delta_{cd}\delta^{cd}\sim N^2 N^2$), as well as one that scales as
 $N^2 g_{YM}^8$ (coming from a contraction that collapses to $\delta_{ab}\delta^{ab}\sim N^2$).
If the above scalings are also valid at strong 't~Hooft coupling and also in general $\mathcal{N}=2$ SCFTs, one would have the following scalings:
\begin{equation}\begin{aligned}
g_{i\bar{j}}&\sim  \lambda^2 \sim N^2  \ ,\\
g_{I\bar{J}}&\sim  \lambda^4 \sim 	N^4 \ ,\\
C_{ijI} &\sim  \lambda^4/N \sim N^3 \ ,\quad (I\sim \textrm{single
  trace}\ \Delta= 4) \ ,\\
C_{ijI} &\sim  \lambda^4 (1+ \frac{1}{N^2}) \sim N^4 + N^2  \ ,\quad (I\sim \textrm{double trace }\ \Delta= 4) \ .
\end{aligned}\end{equation}
Note that we inferred from \eqref{KCH} that on the supergravity side
$g_{i\bar{j}}\sim M^3$, which, using the AdS/CFT dictionary, indeed
implies $g_{i\bar{j}}\sim N^2$ on the dual side.

Putting everything together, the right hand side of (\ref{Riemann}) then scales as
\begin{equation}\begin{aligned}
g_{k\bar{j}}\delta_{i}^{l}+ g_{i\bar{j}}\delta_{k}^{l} &\sim N^2 + N^0 + \dots  \ , \\
\textrm{single trace} &\sim N^3 N^3 N^{-4}N^{-2}  \sim  N^0 +\dots  \ , \\	
\textrm{double trace} &\sim (N^4+N^2) (N^4+N^2) N^{-4} N^{-2} + N^2 \sim N^2 + N^0 + \dots  \ ,
\end{aligned}\end{equation}
Note that the left-hand side of \eqref{Riemann} is independent of $N$
as it is the (scale-invariant) Riemann tensor. This means that at
leading order ($N^2$) the terms on the right-hand side universally
have to cancel each other.\footnote{We thank K.\ Papadodimas for
  extensive discussions on this point.} This predicts, on the one hand,
a certain leading behavior for the OPE coefficients $C_{ijI}$ for
double trace operators.  Moreover, it predicts a
very specific subleading contributions ($N^0$) of the metric
$g_{i\bar{j}}$ and  the double trace OPE coefficients,
as well as a specific leading behaviour  of the OPE coefficients $C_{ijI}$ for
single trace operators. Only if they conspire in the right way, they
can be consistent with
the supergravity result \eqref{eq:modulispace1}.
It would be interesting to check this in explicit SCFTs.

\bibliography{GLSTV}
\providecommand{\href}[2]{#2}\begingroup\raggedright\endgroup

\end{document}